\documentclass[12pt]{article}
\usepackage{a4,epsfig,amsmath,amssymb,graphicx,multirow,oldgerm,euscript,mathrsfs}

\begin{document}

\begin{titlepage}

\begin{flushright}
\begin{tabular}{l}
  PCCF RI 0701\\
  February 2007
\end{tabular}  
\end{flushright}
\renewcommand{\thefootnote}{\fnsymbol{footnote}}

\null\vskip 0.5 true cm

\begin{center}
{{\huge \bf  Testing Time-Reversal: $\boldsymbol{\Lambda_b}$ Decays into Polarized Resonances }}         
\vskip 0.6cm
\vskip 1 true cm
\begin{large}
Z.~J.~Ajaltouni$^{1}$\footnote{ziad@clermont.in2p3.fr}, 
E.~Di Salvo$^{2}$\footnote{Elvio.Disalvo@ge.infn.it}, 
O.~Leitner$^{3}$\footnote{olivier.leitner@lnf.infn.it}
\end{large} 

\bigskip

$^1$ 
Laboratoire de Physique Corpusculaire de Clermont-Ferrand, \\
IN2P3/CNRS Universit\'e Blaise Pascal, 
F-63177 Aubi\`ere Cedex, France\\

\noindent  
$^2$ Dipartimento di Fisica and I.N.F.N. -Sez. Genova,\\
Via Dodecaneso, 33, 16146 Genova, Italy \\  

\noindent  
$^3$ Laboratori Nazionali di Frascati, 
  Via E. Fermi, 40, I-00044 Frascati, Italy \\
  Gruppo teorico, Istituto Nazionale di Fisica Nucleare 
\vskip 1 true cm
\end{center}
\vspace{1.5cm}


\begin{abstract}
Weak decays of beauty baryons like $\Lambda_b$ into $\Lambda  V(J^P=1^-)$, where the 
produced resonances are polarized, offer interesting opportunity to perform tests 
of Time-Reversal Invariance. This paper emphasizes the particular role of the resonance 
polarization-vectors and their physical properties by symmetry transformations. 
In particular, it is shown that the normal component of a polarization-vector, as defined 
in the Jackson's frame, is Lorentz invariant and could get large values, notably in the 
case of  $J/{\psi}$ production.
\end{abstract}
 
\vspace{5.3cm}
PACS Numbers: 11.30.Er, 12.39.-x, 12.39.ki, 13.30.-a, 14.20.Mr

\end{titlepage}

\newpage

\section{Introduction}

An important cornerstone of modern field theories describing the interactions 
among particles is the {\bf CPT} theorem, of  which various proofs have been provided 
by different authors, such as Pauli and Luders. This theorem asserts that 
the product of the three operators {\bf C} (Charge Conjugation), {\bf P} (Parity) 
and {\bf T} (Time Reversal or TR), taken in any order, is a {\it symmetry transformation} 
of the physical laws of Nature. The validity of this theorem is assumed by almost 
all experiments in particle physics.

\noindent 
As an immediate consequence of this theorem, the violation  product of 
two symmetrically operators, {\bf C} and {\bf P}, directly leads to a non conservation 
of the third symmetry operator, {\bf T}. So, the common belief that {\bf T} is violated was 
only based on {\it indirect proof},  until the experiments CP-LEAR and KTeV,  respectively 
performed at CERN and Fermilab,  shed light on  the {\it direct} 
violation of the symmetry {\bf T} in the $K^0 \bar{K^0}$ system. This important fact 
opened the way to search for direct Time Reversal Violation (TRV) in other systems 
of particles: the beauty hadrons, mesons $B$ or beauty baryons like $\Lambda_b$, 
which will be copiously produced  with the forthcoming LHC machine.
\vskip 0.2cm
\noindent
However, direct search for TRV was already suggested by many people, just after the discovery 
of parity breakdown in 1957~\cite{Jackson:1965, Gatto:1958}. The main idea was to look for the 
transverse polarization of the emitted electrons in ${\beta}^-$ decay in neutron or hyperon 
($\Lambda, \Sigma ...$) decays. Transverse polarization of the electron according to the decay 
plane of its parent particle ($n \; \mathrm{or} \; \Lambda ...$) is defined as follows:
$$ {\cal P}_T  =  \langle {\vec s} \cdot {\vec n} \rangle  \;\;  \mathrm{with} \;\;  \vec n = 
\frac{{\vec p_1} \times {\vec p_2}}{|{\vec p_1} \times {\vec p_2}|}\ ,$$
\noindent
where ${\vec p_1} \;  \mathrm{and}  \; {\vec p_2}$ are respectively the momenta of the daughter 
baryon and the emitted electron. $\vec s$ is the electron spin.  It is easy to notice 
that ${\cal P}_T$ changes sign under TR and, assuming TR as a good symmetry, it leads to ${\cal P}_T = 0$.
\newline
Therefore a nonzero ${\cal P}_T$ could be a sign of TRV. In this case, ${\cal P}_T$ is 
called a Time Odd (T-odd) observable. Usually, a T-odd observable 
is not sufficient to prove the violation of TR because of the Final State Interactions (FSI).  It
is also well known that the Final State Interactions during hadronization may modify the 
quantum state of the final particles and, thus could simulate a TRV process. Nevertheless, 
a T-odd physical quantity represents the point of departure to cross-check the Time Reversal 
symmetry, provided an estimation or (in an optimistic case) a complete calculation of the 
FSI are done. In the case where FSI are negligible, a T-odd variable would be a premise of direct TRV.     
\vskip 0.2 cm
\noindent
Since the spin of any particle is Parity-even and T-odd, studying its mean value or its polarization 
provides interesting tools to test both of these two symmetries, especially in weak decay 
processes where parity is known to be violated. It is worth noticing that checking TRV in 
a given decay, like $A  \to  a_1 + a_2$, or in its charge conjugate mode,
$\bar A  \to  {\bar a}_1 + {\bar a}_2$, is not necessarily related to the conservation or 
non-conservation of the {\bf CP} symmetry in these two channels. The method outlined in this 
paper relies on the fact that {\it direct check} of TR symmetry is done without any theoretical assumption
beyond the Standard Model.
\vskip 0.2cm
\noindent
Kinematic calculations are performed in the framework of the Jacob-Wick-Jackson (JWJ)~\cite{JWJ} 
formalism,  while the decay dynamics relies on the OPE techniques supplemented by the Heavy Quark 
Effective Theory (HQET),  in order to do precise estimations of the transition form-factors 
which are involved in the evaluation of the hadronic matrix elements. We refer the reader to 
papers~\cite{Leitner:2006nb,OLZJA:2006,Ajaltouni:2004zu} for all calculation details.

\noindent     
The aim of the present paper is to stress the particular role of the polarization-vectors of 
both the hyperon $\Lambda$ and the vector-meson $V(J^P=1^-) = J/\psi, \rho^0, \omega$ coming 
from $\Lambda_b$ decays. Section II is devoted to set helicity states as well as cross-section 
in $\Lambda_b$ baryon decays. The component calculations and the  behavior of the polarization-vectors 
according to the $\Lambda_b$ initial polarization density matrix (PDM) are analyzed in Section III. 
The use of polarization-vectors on search for TR violation is also discussed in this section. 
In Section IV, emphasis is put on the choice of some special frames which help to clearly exhibit 
the T-odd observables related to the resonance polarizations. Finally, results are discussed and 
a few conclusions are drawn in Section V.

\section{ Production and decay of $\boldsymbol{\Lambda_b}$ baryon}

An essential point of our study is based on assuming that the $\Lambda_b$ baryons which 
are produced in $p-p$  collisions are transversely polarized like the ordinary 
hyperons $(\Lambda, \Sigma, ...)$ in hadron-hadron or hadron-nucleus collisions. Because 
of the $\Lambda_b$ spin $1/2$, its Polarization Density Matrix (PDM) is a $(2 \times 2)$ hermitian 
matrix whose trace, $Tr({\rho}^{\Lambda_b})$, is normalized to one. The only complex elements are the 
non-diagonal ones,  which verify ${\rho}^{\Lambda_b}_{12} = {{\rho}^{\Lambda_b}_{21}}^*$. No more 
assumptions are made concerning this density matrix. Moreover, the angular momentum states are 
constrained by the conservation of the total angular momentum: $\vec {s_{\Lambda_b}} = \vec {s_{\Lambda}}  
+ \vec {s_V} + \vec L  \; \;  \mathrm{with} \; \; {s_{\Lambda_b}}={s_{\Lambda}} = 1/2\ , \; s_V = 1 \;\;  
\mathrm{and} \; \; \vec L$ is the orbital angular momentum in the $\Lambda_b$ rest-frame. Taking the 
direction of $\vec {p_{\Lambda}} = (p, \theta, \phi)$ as the quantization axis one, the contribution 
of the orbital angular momentum $\vec L$ will be suppressed and only four helicity states 
remain to describe the $\Lambda V$ system:
$$ ({\lambda}_{\Lambda}, {\lambda}_V ) = (1/2,0), \;  (-1/2, -1), \; (1/2, 1), \; (-1/2, 0)\ , $$ 
\noindent
to which correspond the hadronic matrix elements 
$\mathcal {A}_{(\lambda_1,\lambda_2)}(\Lambda_b \to \Lambda V)$.
Taking into account the ${\Lambda_b}$ PDM, the cross-section of the $\Lambda_b  \to  \Lambda V(1^-) $ decay  
is therefore given by the following relation: 

\begin{eqnarray}
d\sigma \propto  \sum_{M_i,M^{\prime}_i} {\sum_{\lambda_1, \lambda_2}}{\rho}_{M_i M^{\prime}_i}^{\Lambda_b}
{|\mathcal {A}_{(\lambda_1,\lambda_2)}(\Lambda_b \to \Lambda V)|}^2 d_{M_i {\lambda}}^{1/2} 
d_{M^{\prime}_i {\lambda}}^{1/2}
{\exp{i(M^{\prime}_i - M_i)\phi}}\ . 
\end{eqnarray} 
     
\noindent
Then, we introduce the helicity asymmetry parameter, ${\alpha}_{AS}$, which is the 
analogous of the $\Lambda$ asymmetry one in the standard decay $\Lambda \to p {\pi}^-$, 
 
\begin{eqnarray}
 \alpha_{AS} = \frac{{|{\Lambda_b}(+)|}^2-{|{\Lambda_b}(-)|}^2}{{|{\Lambda_b}(+)|}^2 
+{|{\Lambda_b}(-)|}^2}\ ,
\end{eqnarray}
with the elements $\Lambda_b(\pm)$ defined by,
\begin{align}
{|{\Lambda_b}(+)|}^2 = &{|\mathcal {A}_{(1/2,0)}(\Lambda_b \to \Lambda V)|}^2  
+{|\mathcal {A}_{(-1/2,-1)}(\Lambda_b \to \Lambda V)|}^2\ , 
\nonumber \\
{|{\Lambda_b}(-)|}^2 = &{|\mathcal {A}_{(-1/2,0)}(\Lambda_b \to \Lambda V)|}^2 
+{|\mathcal {A}_{(1/2,1)}(\Lambda_b \to \Lambda V)|}^2\ . 
\end{align}
Therefore, the differential cross-section gets the following form:
\begin{eqnarray}\label{qe01}
\frac{d\sigma}{d\Omega} \propto  1 +  {\alpha_{AS}}{{\cal P}^{\Lambda_b}}{\cos \theta} 
+ 2{\alpha_{AS}}{\Re e{({\rho}_{+-}^{\Lambda_b} \exp{i\phi})}} {\sin \theta}\ .
\end{eqnarray}
\noindent
In Eq.~(4), ${\cal P}^{\Lambda_b} (= {\rho}^{\Lambda_b}_{++} - {\rho}^{\Lambda_b}_{--})$ is the 
value of the $\Lambda_b$ polarization. The above relations show the importance 
of the $\Lambda_b$ polarization and its PDM  in the angular distributions of 
the resonances, $\Lambda \; \mathrm{or} \; V$, in the $\Lambda_b$ rest-frame. 

\section{Polarizations of the final resonances}

Basic principles of Quantum Mechanics allow us to deduce the spin density 
matrix of the final $\Lambda V$ system,  which is an essential parameter to compute 
the polarization-vector of each resonance, ${\rho}^f = \mathcal{T}^{\dagger}  
{\rho}^{\Lambda_b}\mathcal{T}$. $\mathcal{T}$ is the transition-matrix 
(the $\mathcal{S}$-matrix being defined by 
$\mathcal{S} = 1 + i\mathcal{T}$) whose elements are explicitly given  in Ref.~\cite{Leitner:2006nb}. The 
normalization of the matrix ${\rho}^f $ is obtained by $Tr{({\rho}^f)}  
=  \frac{d\sigma}{d\Omega} =  N  W(\theta,\phi)$,
where $Tr$ is the trace operator and $N$ is a normalization constant. Consequently, 
the polarization-vector of any resonance $R_{(i)} \ ( R_1 = \Lambda, \ R_2 = V) $ is defined
by
\begin{eqnarray}
\vec{\mathcal{P}_i}  =   \langle \vec{S_i} \rangle  =   \frac{Tr{\big( {\rho}^f_i {\vec S_i} 
\big)}}{Tr({\rho}^f_i)}\ , 
\end{eqnarray}
where ${\rho}^f_i$ is the spin density-matrix of the resonance, $R_{(i)}$, deduced from ${\rho}^f$.
As the final state is a {\it composite system} made out of two particles with different spin ($s_1 =
1/2, s_2 = 1$), each ${\rho}^f_i$ will be obtained from the general expression of 
 ${\rho}^f$ by summing up over the degrees of freedom of the other resonance. Thanks 
to this method, we can obtain ${\rho}^{\Lambda}\  \mathrm{and} \  {\rho}^V$. See Ref.~\cite{Leitner:2006nb}
 for all analytical results.

\subsection{Choice of particular frames}
In the $\Lambda_b$ rest-frame\footnote{The $\Lambda_b$ rest frame used in our analysis is given in Figure 1.}, 
a specific frame according to the Jackson's method is 
constructed for each resonance $R_{(i)}$ (the index $i$ will be dropped for convenience): 
\begin{eqnarray}\label{qe03}
 \vec{e}_L =  \frac{\vec{p}}{p}\ , \ \ \    
 \vec{e}_T =  \frac{\vec{e}_Z \times \vec {e}_L}{|\vec{e}_Z \times \vec{e}_L|}\ , \ \ \   
 \vec{e}_N = \vec{e}_L \times \vec{e}_T\ , 
\end{eqnarray}
\noindent
where $\vec {e_Z} \; \mathrm {is \ parallel \ to}  \; \vec n$. $\vec n$ 
is initially defined as the normal unit-vector to the $\Lambda_b$ production plane. Then, 
each polarization-vector can be expressed as:
\begin{eqnarray}
\vec{\mathcal{P}} = {P_L} \vec{e}_L + {P_N} \vec{e}_N + {P_T}  \vec{e}_T\ , 
\end{eqnarray}
\noindent 
where $P_L, P_N \ \mathrm{and} \ P_T$ are respectively the 
{\it longitudinal, normal and transverse} components of $\vec {\cal P}$. It is worth 
noticing that the basis vectors $\vec {e_L}, \vec {e_T} \;  \mathrm{and} \; \vec {e_N} $
have the following properties according to parity and TR: P-odd,T-odd;  P-odd,T-odd  and  P-even, 
T-even respectively, while the polarization-vector $\vec {\cal P}$ is P-even and T-odd. So, any component 
of $\vec {\cal P}$ defined by the scalar product
 $P_j = \vec{\mathcal{P}} \cdot \vec {e}_j \;  \mathrm{with} \; j = L,N,T$ gets transformed 
as:
$$  P_L = \mathrm{P-odd, T-even} , \;  P_T = \mathrm{P-odd, T-even} \; \mathrm{and}  \; P_N =
\mathrm{P-even, T-odd}. $$
\noindent
Note that the longitudinal axis defined 
by $\vec {e_L}$ is taken as the quantization axis. $\vec {e_N} \; \mathrm{and} \;  \vec {e_T}$ 
are identified to $x \; \mathrm{and} \; y$ axis, respectively. The previous relation allows us to 
write down the formal expression of any $\vec {\cal P}_i$ by expanding the trace operator 
over the different spin states\footnote{In order to perform these calculations, some simple 
and fundamental relations are used, whatever the spin is:
$$ S_x |{\lambda}\rangle = {(|{\lambda}+1\rangle + |{\lambda}-1\rangle)}/{\sqrt 2}\ , \\
  S_y |{\lambda}\rangle = {i(-|{\lambda}+1\rangle +|{\lambda}-1\rangle)}/{\sqrt 2}\ , \\
    S_z|{\lambda}\rangle = {\lambda}|{\lambda}\rangle\ . $$

\noindent
(Letters indicating other physical parameters are dropped for simplicity). Technical 
details of the computation of both $\vec {{\cal P}^{\Lambda}} \; \mathrm{and} \; 
\vec {{\cal P}^V}$ are given in Ref.~\cite{Leitner:2006nb}.} and  one gets,
\begin{eqnarray}  
\vec{\mathcal{P}}_i  \; W(\theta,\phi)  =   N \; {\Sigma}_{\lambda} \Bigl(\langle\theta,\phi,
{\lambda}|{\rho}^f_i {\vec S}|\theta,\phi,{\lambda}\rangle \Bigr)\ ,
\end{eqnarray}
 $W(\theta,\phi)$ being defined at the beginning of Section 3. 

\subsection{$\boldsymbol{\Lambda}$ and $\boldsymbol{V(J^P=1^-)}$ polarization-vectors}
The matrix elements given in the right-handed side of Eq.~(8) can be explicitly 
calculated~\cite{Jackson:1965} and the three components of $\vec{\mathcal{P}}^{\Lambda}$ 
get the following expressions:
\begin{align}
P_x^{\Lambda} \ {W(\theta,\phi)} \  & \propto  \  2 \Re e{(\langle\theta,\phi,1/2|{\rho}^{\Lambda}| 
\theta,\phi,-1/2 \rangle)}\ , \nonumber  \\
P_y^{\Lambda} \ {W(\theta,\phi)} \  & \propto  \ -2 \Im m{(\langle \theta,\phi,1/2|{\rho}^{\Lambda}| 
\theta,\phi,-1/2 \rangle)}\ , \nonumber \\
P_z^{\Lambda} \ {W(\theta,\phi)} \  & \propto  \  {\bar \omega}(+1/2) - {\bar \omega}(-1/2)\ ,   
\end{align}
where the ${\bar \omega}(\pm)$ are defined in~\cite{Leitner:2006nb}. The vector meson has 
spin $S_{(2)}= 1$ and therefore three helicity states. Based on
\begin{eqnarray}
\vec{\mathcal{P}}^V  \ W(\theta,\phi)  =   N \
{\Sigma}_{\lambda_2} \Bigl({\Sigma}_{\lambda_1}\langle\theta,\phi,{\lambda_1},{\lambda_2}|{\rho}^f {\vec
S}|\theta,\phi,{\lambda_1},{\lambda_2}\rangle \Bigr)\ ,
\end{eqnarray}
the components of  $\vec {{\cal P}^V}$ are obtained in the same manner, although more tedious. One has 
\begin{align}
P_x^V \ {W(\theta,\phi)} \  & \propto  \  {\sqrt 2} \Re e{\big((\langle 0|{\rho}^V|-1 \rangle) + (\langle
1|{\rho}^V|0 \rangle)\big)} \ , \nonumber  \\
P_y^V \ {W(\theta,\phi)} \  & \propto  \  {\sqrt 2} \Im m{\big((\langle 0|{\rho}^V|-1 \rangle) + (\langle
1|{\rho}^V|0 \rangle)\big)} \ , \nonumber  \\
P_z^V \ {W(\theta,\phi)} \  & \propto \  (\langle 1|{\rho}^V|1 \rangle) - (\langle -1|{\rho}^V|-1 \rangle)\ .
\end{align}
\noindent
As it was expected, the helicity value ${\lambda}_V = 0$ does not contribute to the longitudinal
polarization of the vector-meson. We also underline the importance of the initial $\Lambda_b$ 
polarization, $\mathcal{P}^{\Lambda_b}$, as well as 
the non-diagonal matrix element, ${\rho}_{+-}^{\Lambda_b}$ in the components of $\vec{\mathcal{P}}^V$. 
Details can be found in Ref.~\cite{Leitner:2006nb}.

On the dynamical side, the Heavy Quark Effective Theory~\cite{Korner:1994nh, Hussain:1994zr,
Neubert:1993mb} (HQET) formalism is used to evaluate the hadronic form factors
involved in $\Lambda_b$-decay. Weak transitions including heavy quarks can be
safely described when the mass of a heavy quark is large enough compared to the
QCD scale, $\Lambda_{QCD}$. Properties such as flavor and spin symmetries can
be exploited in such a way that corrections of the order of $1/m_Q$ are systematically
calculated within an effective field theory. Then, the hadronic amplitude of the
weak decay is investigated by means of the  effective Hamiltonian, $\Delta B=1$,
where the Operator Product Expansion formalism separates the soft and hard regimes. All 
results about transition form factors as well as hadronic matrix elements are 
given in Ref.~\cite{Leitner:2006nb}.

\subsection{Numerical results}
$\bullet$ In a first step, values for all input parameters are taken 
from~\cite{Leitner:2006nb}. The initial $\Lambda_b$ polarization and $\Lambda_b$ 
polarization density matrix element used in our numerical computations are   
${\cal P}^{\Lambda_b} =  100\% \; \mathrm{and} \;  {\Re e({\rho}_{+-}^{\Lambda_b})}  
= {\Im m({\rho}_{+-}^{\Lambda_b})} =   {\sqrt{2}}/2$, respectively. The corresponding 
spectra of the three components of ${\vec{\cal P}}^{\Lambda} \; \mathrm{and} \; 
  {\vec{\cal P}}^V $ are shown in Figures~2 and~3. Comments on longitudinal, 
transverse and normal components of these polarization vectors are the following: 
(i) the longitudinal components, $P_L = P_z$, of both the two resonances are asymmetric 
because of {\it parity violation} in weak $\Lambda_b$ decays; (ii) the spectra of 
the transverse components, $P_T = P_y$, are quite symmetric and their asymmetries  
are $\approx 1.0\%$; (iii) the normal components, $P_N = P_x$, are clearly 
asymmetric. Their asymmetry values are respectively $23\% \; \mathrm{and} \; -54\%$ 
for $\Lambda$ and $J/{\psi}$. 
\vskip 0.2cm
\noindent
$\bullet$ In a second step, attempts to understand correlations between  the $\Lambda_b$ 
initial polarization and the physical properties of its decay products are made. All 
results are obtained with Monte-Carlo simulations by varying independently ${\cal P}^{\Lambda_b} 
\ \mathrm{and} \ {\rho}_{+-}^{\Lambda_b} $. Non-diagonal matrix
elements being generally unknown, we set ${\Re e({\rho}_{+-}^{\Lambda_b})}  
= {\Im m({\rho}_{+-}^{\Lambda_b})} =  0$ and we let ${\cal P}^{\Lambda_b}$ vary between $100\% \
\mathrm{and} \ 0\%$. The resulting spectra of the normal components, $P_N^{\Lambda} \ \mathrm{and} \
P_N^{J/{\psi}} $ are usually {\it sharp}, while the transverse components, both 
$P_T^{\Lambda} \ \mathrm{and} \ P_T^{J/{\psi}}$, are {\it always equal to zero}. These two physical
properties can be explained as direct consequences of ${\rho}_{+-}^{\Lambda_b} = 0$. \\
\noindent
In Table 1, mean values and asymmetry parameters of the normal component spectra for  $\Lambda$ 
and $J/{\psi}$ are respectively listed. Interesting remarks can be drawn: (i) for ${\cal P}^{\Lambda_b} 
\neq  0$,  the normal components are largely dominating and their asymmetries are nearly 
equal to minus one; (ii) for ${\cal P}^{\Lambda_b} =  0$, the $J/{\psi}$ normal component is still 
dominating ($\approx -0.8 $) while  $P_N^{\Lambda}$ is equal to zero, the 
${\Lambda}$ polarization being completely longitudinal ($P_L^{\Lambda} = -100\% $).
\vskip 0.2cm
\noindent

In order to understand the role of the $\Lambda$ azimuthal distribution and its 
effects on the resonance polarization-vectors, a comparison between two series of $P_N$ 
spectrum is performed. One series  is obtained with ${\Re e({\rho}_{+-}^{\Lambda_b})} 
= {\Im m({\rho}_{+-}^{\Lambda_b})} = 0 \ $, while the other one is obtained with the standard values, 
${\Re e({\rho}_{+-}^{\Lambda_b})} = {\Im m({\rho}_{+-}^{\Lambda_b})} =   {\sqrt{2}}/2$. In Figures 4 and 
5, the spectra of $P_N$ and $P_T$ for $\Lambda$ and $J/{\psi}$ are respectively plotted. One
notes that the spectra belonging to ${\Re e({\rho}_{+-}^{\Lambda_b})} = {\Im m({\rho}_{+-}^{\Lambda_b})} 
\neq 0$ are much broader than those belonging to ${\rho}_{+-}^{\Lambda_b} = 0$. Moreover, the absolute 
values of the asymmetry parameters decrease for both $P_N^{\Lambda} \ \mathrm{and} \ P_N^{J/{\psi}}$ 
when ${\rho}_{+-}^{\Lambda_b} \neq  0 \ $, while the transverse components remain symmetric with a mean 
values around $0$.  
\noindent
Whatever the $\Lambda_b$ PDM elements are, this exhaustive study indicates that 
the normal components of the polarization-vectors, which are T-odd observables, must be taken as 
serious candidates to cross-check TR symmetry.

\section{Specific calculation of the normal component $\boldsymbol{P_N}$}

\subsection{Relativistic form}

In the previous paragraph, the resonance polarization-vectors are estimated by standard 
non-relativistic quantum mechanical methods in the $\Lambda_b$ rest-frame. However, 
in order to measure ${\vec {\cal P}^{(i)}}$ for each resonance, $R_{(i)}$, and its 
transformation by symmetry operations like Parity and Time-Reversal, a thorough 
examination of the $R_{(i)}$ decay products by sophisticated methods (Byers, Dalitz)~\cite{Leader:2001}
 must be done in the resonance rest-frame itself. 
These calculations will not be developed in 
the present paper, but a rigorous method using the relativistic spin will be applied in order 
to understand the modification of ${\vec {\cal P}^{(i)}}$ and, essentially its normal component $P_N$, 
when considering two different rest-frames: the $\Lambda_b$ one and the  $R_{(i)}$ resonance one,  
which are related by a Lorentz transformation.
\noindent
Thus, we are led to study the spin of any particle and its polarization in their {\it relativistic form}.
The latter is described by an axial four-vector $(S^{\mu}) = (S^0, \vec S )$ which verifies the 
fundamental relation: 
$p_{\mu} S^{\mu} = 0\ , \; p_{\mu} $ being the 4-momentum of the particle. 
\noindent
In what follows, we will drop the index $(i)$ which designates any resonance ($\Lambda \; \mathrm{or} \;
 J/{\psi}$) and we will assign a prime to the physical quantities defined in the $R_{(i)}$ rest-frame.
\vskip 0.2cm
\noindent

\subsection{Different rest-frames}

${\cal P}^\prime =  (0, \vec {P ^\prime})$ is the polarization 4-vector of the 
$\Lambda$ hyperon in its own rest-frame. Taking the  $\Lambda$ hyperon at rest, the beauty baryon $\Lambda_b$ 
will move with a momentum $\vec {p^\prime}$ (with respect to $\Lambda$) and an energy 
$E' = \sqrt{{p'}^2 + M^2} \; \mathrm{where}  \;  M  \; \mathrm{is \; the \; } \Lambda_b$ mass. 
In the following, we set $ \cosh {\alpha} = {E'/M} \; \mathrm{and} \; \sinh{\alpha} = {p'/M}$. The 
unit-vector ${\vec e}_N$ previously defined  in the $\Lambda_b$ rest-frame can be decomposed like: \\
$$ {\vec e}_N = \vec {e_{N \parallel}} + \vec {e_{N \perp}} \; \;  \mathrm{with} \;  \;  \vec {e_{N \parallel}}=
\frac{({\vec {e_N}} \cdot {\vec p})}{p^2} {\vec p} \; \; \mathrm{and} \; \; {\vec e}_{N \perp} = {\vec e}_N -
{\vec e}_{N \parallel}\ , $$ 
\noindent
where $\vec p$ is the $\Lambda$ momentum in the $\Lambda_b$
rest-frame. In order to get the transformed of $\vec {e_N}$  in the $\Lambda$ rest-frame, $\vec {e^\prime_N}$, a 
new 4-vector must be defined in the $\Lambda_b$ rest-frame: $e_N = (0, \vec {e_N}) = (0,\vec {e_{N \parallel}}) 
+ (0, \vec {e_{N \perp}})$. It is worth noting some important kinematic properties: (i)
$\vec {e_N}$ is {\it orthogonal} to $\vec {p_{\Lambda}} = \vec p$ in the $\Lambda_b$ rest-frame, 
${\vec e}_{N \parallel} = \vec 0,  \mathrm{which \ implies} \; {\vec e}_{N \perp} = {\vec e}_N$; (ii)
the momenta $\vec p \; \mathrm{and} \; \vec {p^\prime}$ are parallel. Therefore the 
corresponding transformations of the 4-vector $e_N \ \mathrm{into} \ e^{\prime}_N$ are the following:  
(a) $e^\prime_{N \perp} = (0, {\vec {e^\prime}}_{N \perp}) = (0, {\vec e}_{N \perp})$, because  
the orthogonal component to the boost-direction is unmodified; (b) $e^\prime_{N \parallel} 
= (\sinh{\alpha} \; |{\vec e}_{N \parallel}| , \cosh{\alpha} \;  {\vec e}_{N \parallel}) \; 
\mathrm{with} \;  {\vec e}_{N \parallel} = \vec 0$. The final expression of the 4-vector $e^\prime_N$ 
will be given by
$$ e^\prime_N 
= e^\prime_{N \parallel} + e^\prime_{N \perp} \Rightarrow  e^\prime_N = (0, {\vec e}_{N \perp}) = e_N\ .$$ 
\noindent
Using the notations above, it can be noticed that, in each rest-frame, the normal component
of the polarization-vector is given by: 
\begin{align}
\mathrm{\Lambda \ rest-frame}  \;  \; &: {\cal P}^\prime \cdot {e^\prime}_N = - {\vec P}^\prime \cdot
{\vec {e^\prime}}_N = -{P^\prime}_N\ , \nonumber \\
\mathrm{\Lambda_b \ rest-frame} \;  \; &:  {\cal P} \cdot {e_N} = - {\vec P} \cdot {\vec e}_N = -P_N\ .
\end{align}
\noindent
By Lorentz-Invariance, one can easily deduce that ${\cal P} \cdot {e_N} =  {\cal P}^\prime \cdot {e^\prime}_N$ 
which obviously gives $P_N = {P^\prime}_N. $ So, the normal-component of the polarization-vector as 
defined in the previous frame, ${({\vec e}_L, {\vec
e}_T, {\vec e}_N )}$, is a {\it Lorentz Invariant}. This interesting physical property allows us to
cross-check TR symmetry either in the $\Lambda_b$ rest-frame or in any resonance rest-frame coming from
$\Lambda_b$ decays. 

\newpage
\section{Conclusion} 
Complete calculations based both on the helicity formalism (kinematics) and on the OPE techniques
supplemented by HQET (dynamics) have been performed in a rigorous way for a precise determination of the 
physical properties of the $\Lambda_b \to  \Lambda V(J^P=1^-)$ decays. Resonances $\Lambda \ \mathrm{and} \
V(J^P=1^-)$ being polarized, it is shown that the normal components of their polarization-vectors are T-odd
observables. Furthermore, these components have large asymmetries and they are Lorentz-invariant. An exhaustive
study of $P_N^{\Lambda, J/\psi} $ has been performed according to the $\Lambda_b$ polarization density
matrix. Thanks to our analysis, it is confirmed that these observables are truly serious candidates 
to cross-check Time-Reversal symmetry, and we hope to detect these effects with the forthcoming LHC machine.

\subsection*{Acknowledgments}
Thanks to a meeting held at the International QCD Conference, QCD06, in Montpellier, 
two of the authors, E.D.S. and Z.J.A., began their discussion about the relativistic 
polarization. They would like to thank very warmly the Chairman of the organizing committee, 
S. Narison, for the opportunity he gave them at this meeting.  One of us (Z.J.A.) would 
like also to thank Dr Vincent Morenas for illuminating discussions concerning the polarization issue.

\newpage
{\large
\begin{table}[htp]
\begin{center}
\begin{tabular}{|c|cc|cc|}
\hline
${\cal P}^{\Lambda_b}$&$P_N^{\Lambda}$&$As^{\Lambda}$&$P_N^{J/\psi}$&$As^{J/\psi}$\\
\hline
\hline
$100\%$&-0.98&-1.0&-0.88&-0.95\\
\hline
$75\%$&-0.97&-1.0&-0.89&-1.0\\
\hline
$50\%$&-0.96&-1.0&-0.87&-1.0\\
\hline
$25\%$&-0.88&-1.0&-0.85&-1.0\\
\hline
$10\%$&-0.61&-1.0&-0.83&-1.0\\
\hline
$0\%$&0.0&0.0&-0.81&-1.0\\
\hline
\end{tabular}
\end{center}
\caption{Mean values, $P_N^{\Lambda,J/\psi}$, and asymmetries, $As^{\Lambda, J/\psi}$, of 
the polarization-vector normal components of $\Lambda \ \mathrm{and} \ J/{\psi}$, 
respectively\label{tab1}. Results are given as functions of 
the initial $\Lambda_b $ polarization varying from $100\%$ to $0\%$ and for the 
$\Lambda_b \to \Lambda J/{\psi}$ decay channel.}
\end{table}
}
\vskip 3.0cm
\begin{figure}[hpt]
\begin{center}
\mbox{\epsfig{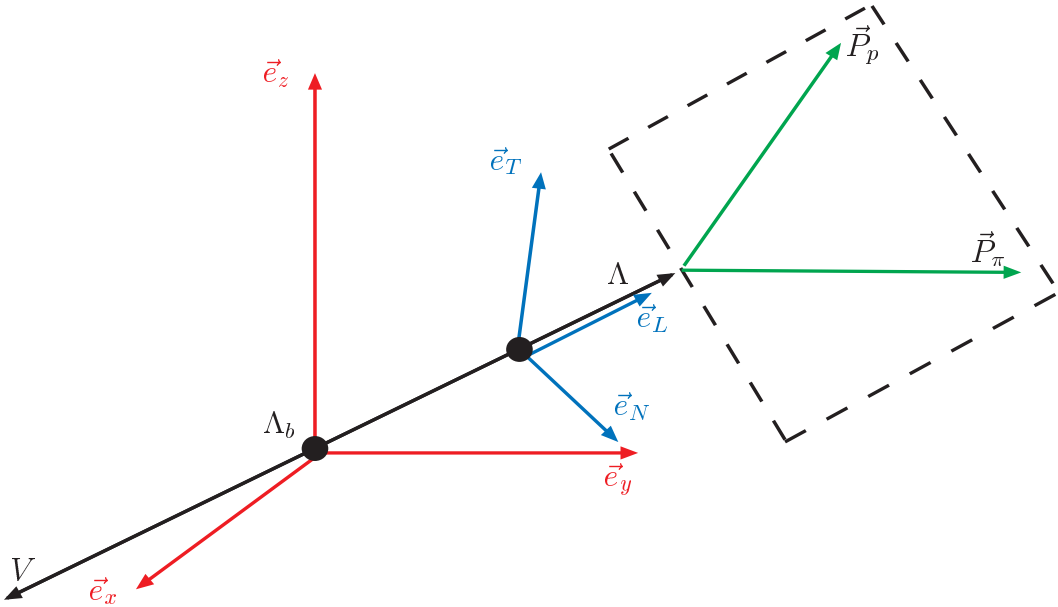}}
\end{center}
\caption{The $\vec e_x, \vec e_y, \vec e_z$ as well as the $\vec e_T, \vec e_N, 
\vec e_L$ frames in the $\Lambda_b$ rest-frame.} 
\end{figure}

\begin{figure}[hpt]
\begin{center}
\mbox{\epsfig{file = 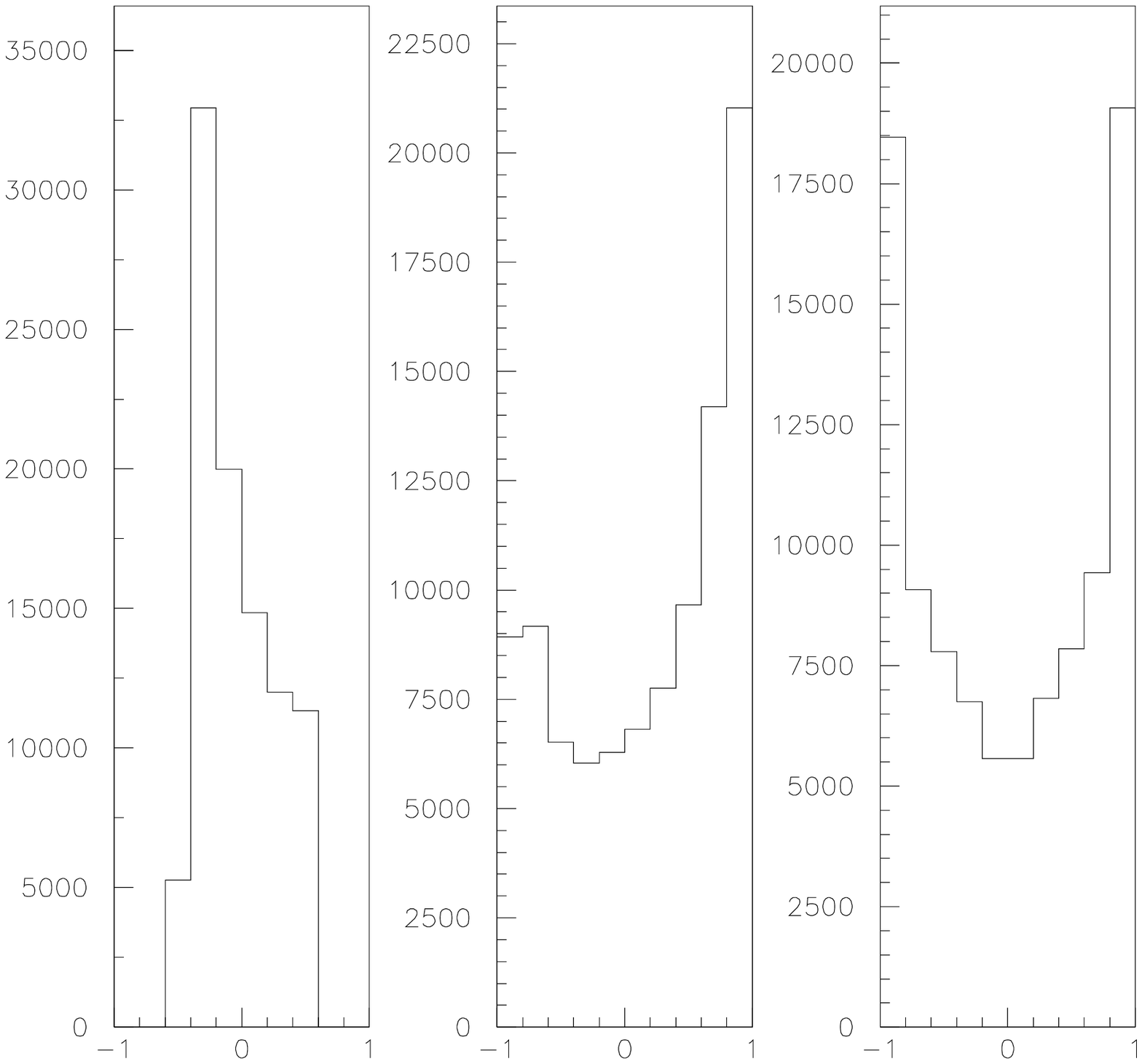, height = 9.0cm, width = 16.0cm}}
\end{center}
\caption{Spectra of the $\Lambda$ polarization-vector components: (from 
left to right) $P_L, P_N, P_T$, respectively in case of  ${\cal P}^{\Lambda_b} 
=  100\% \; \mathrm{and} \;  {\Re e({\rho}_{+-}^{\Lambda_b})}  
= {\Im m({\rho}_{+-}^{\Lambda_b})} =   {\sqrt{2}}/2$.} 
\end{figure}
\begin{figure}[hpt]
\begin{center}
\mbox{\epsfig{file = 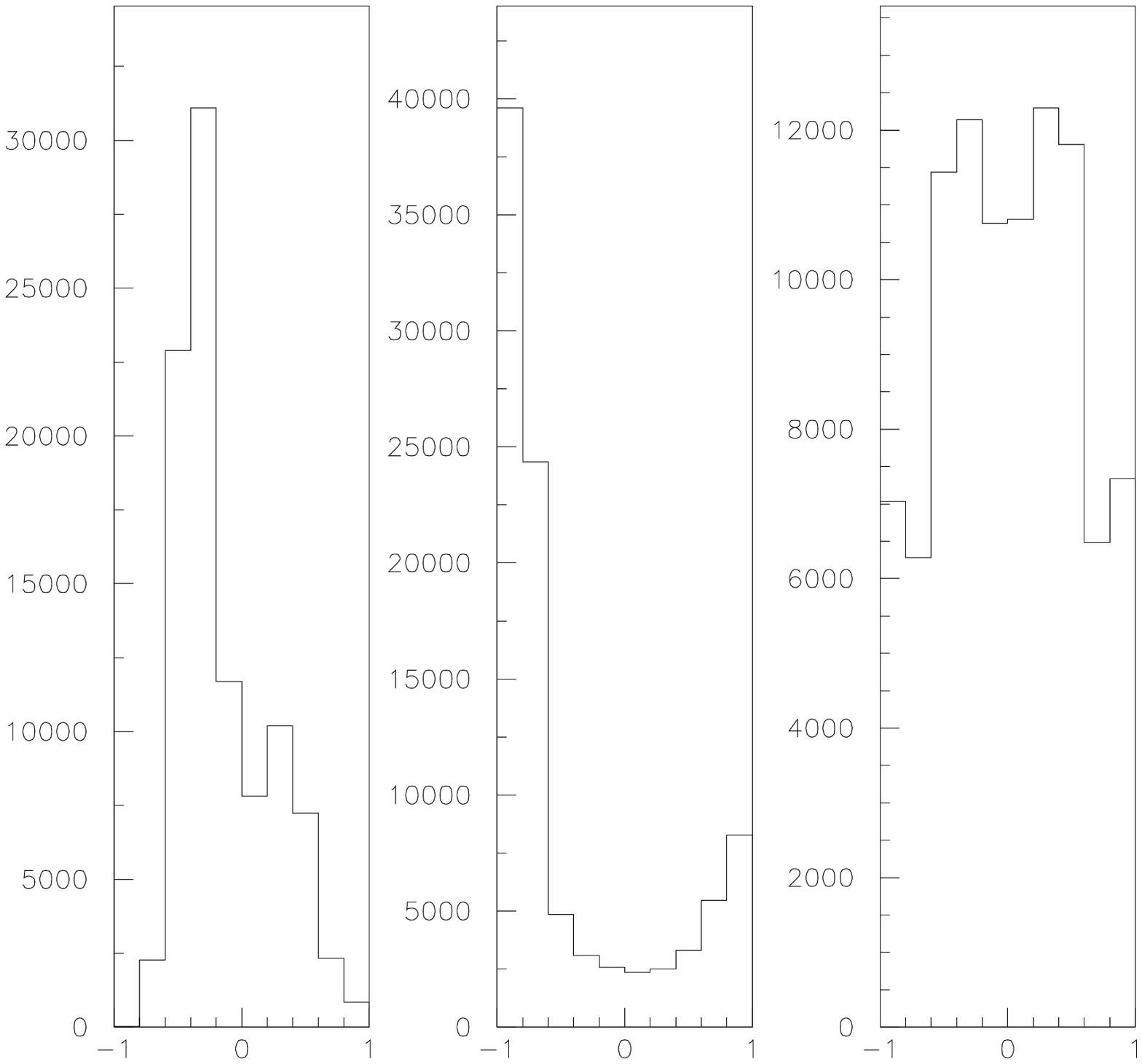, height = 9.0cm, width = 16.0cm}}
\end{center}
\caption{Spectra of the $J/{\psi}$ polarization-vector components: (from 
left to right) $P_L, P_N, P_T$, respectively in case of ${\cal P}^{\Lambda_b} 
=  100\% \; \mathrm{and} \;  {\Re e({\rho}_{+-}^{\Lambda_b})}  
= {\Im m({\rho}_{+-}^{\Lambda_b})} =   {\sqrt{2}}/2$.} 
\end{figure}
\begin{figure}[hpt]
\begin{center}
\mbox{\epsfig{file = 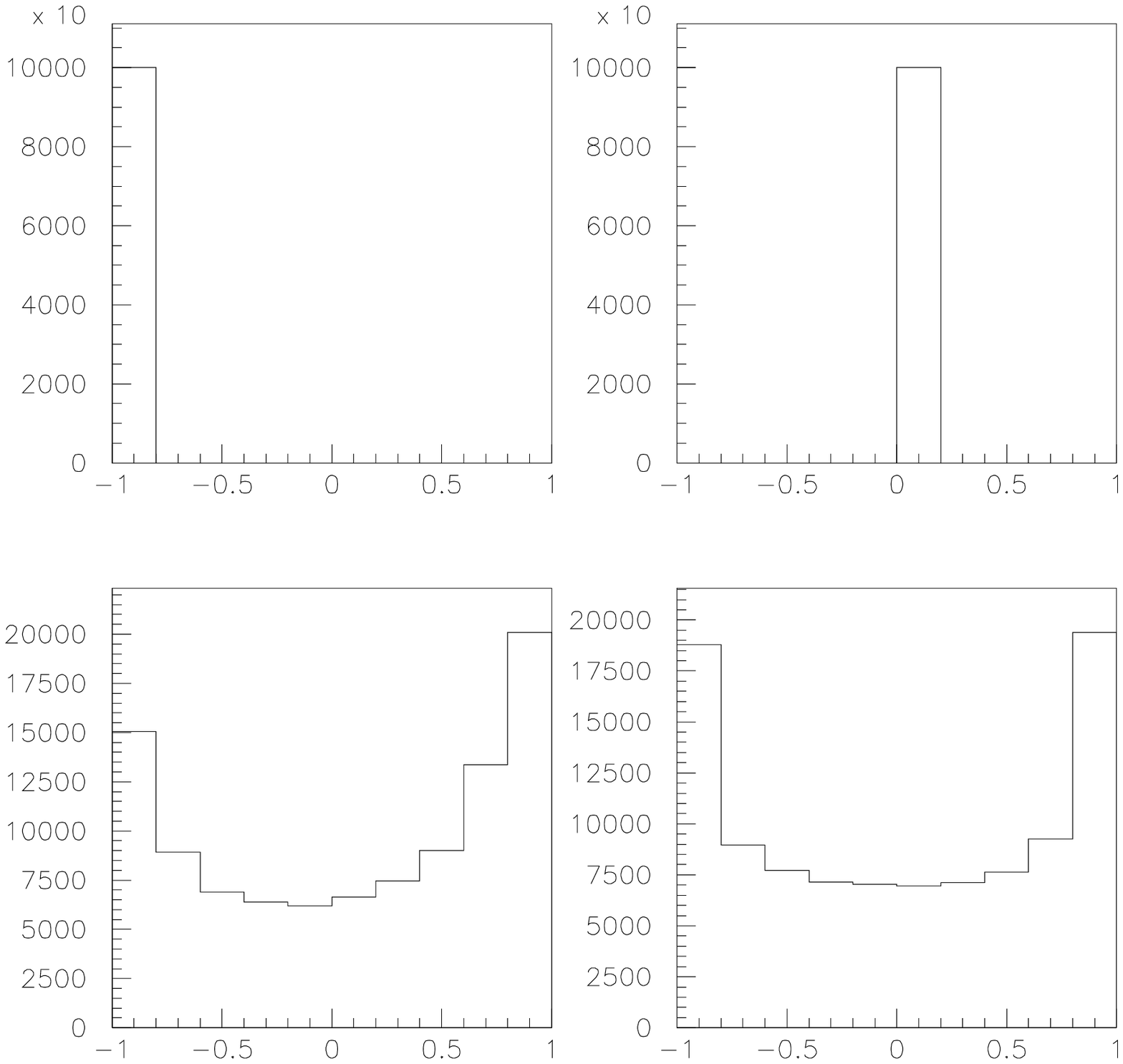, height = 9.0cm, width = 16.0cm}}
\end{center}
\caption{Spectra of $P_N^{\Lambda} \ \mathrm{and} \  P_T^{\Lambda} \ \mathrm{with} 
\ {\cal P}^{\Lambda_b} =50\%$. Upper histograms correspond to the case of 
${\Re e({\rho}_{+-}^{\Lambda_b})} = {\Im m({\rho}_{+-}^{\Lambda_b})}= 0$,  while lower 
histograms correspond to ${\sqrt 2}/2$.}  
\end{figure}
\begin{figure}[hpt]
\begin{center}
\mbox{\epsfig{file = 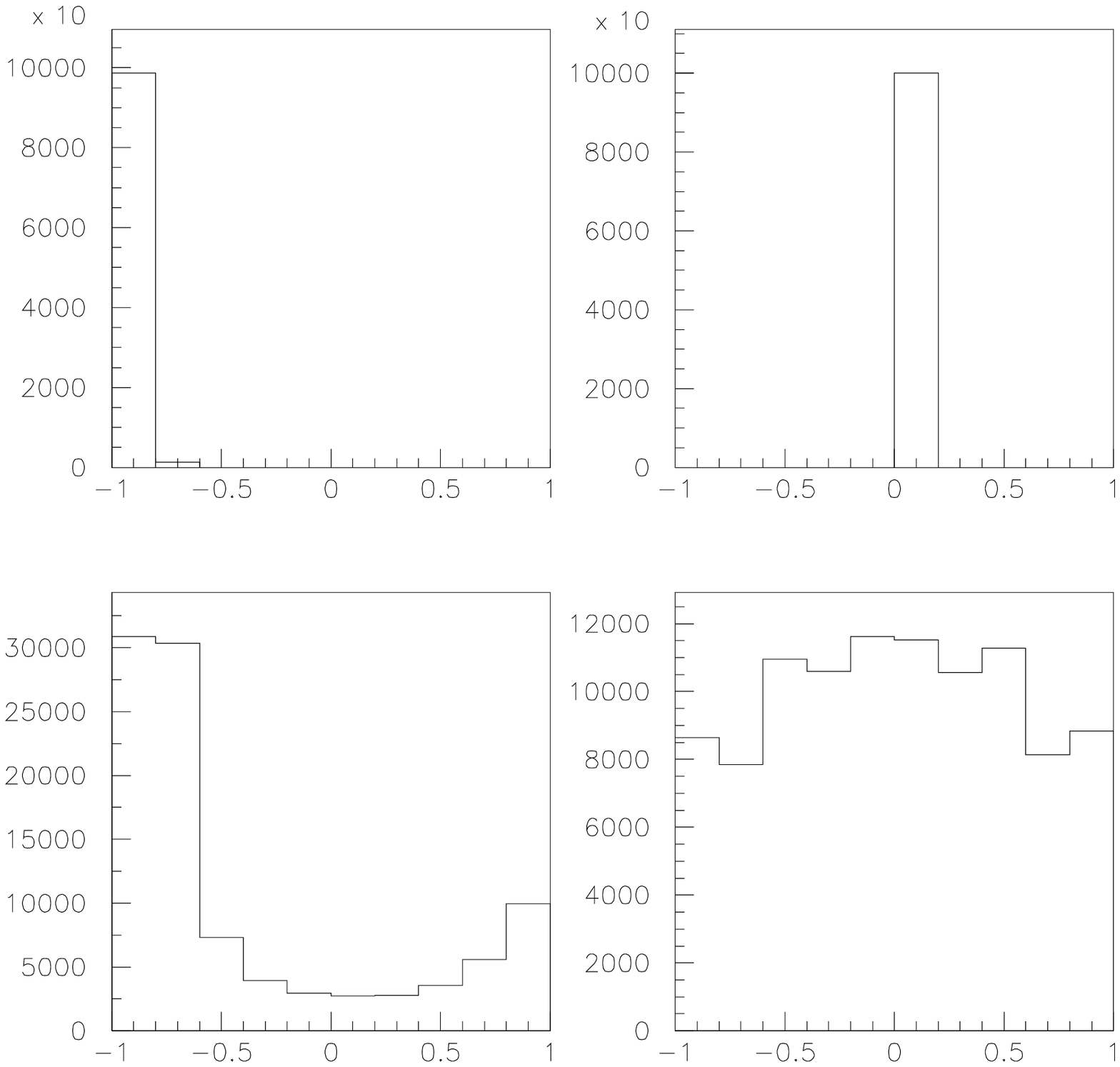, height = 9.0cm, width = 16.0cm}}
\end{center}
\caption{Spectra of $P_N^{J/{\psi}} \ \mathrm{and} \  P_T^{J/{\psi}} \ \mathrm{with} 
\ {\cal P}^{\Lambda_b} =50\%$. Upper histograms correspond to the case of 
${\Re e({\rho}_{+-}^{\Lambda_b})} = {\Im m({\rho}_{+-}^{\Lambda_b})} = 0$, while lower 
histograms correspond to ${\sqrt 2}/2$.}  
\end{figure}

\end{document}